\documentstyle [12pt,epsf]{article}

\input{epsf}
\begin{document}

\begin{titlepage}
\begin{center}
\vspace{-0.1in}

{\large \bf Path integral approach \\ to \\the full Dicke model}\\
 \vspace{.3in}{\large\em M.~Aparicio Alcalde,\footnotemark[1] and
 B. M. Pimentel\,\footnotemark[2]}\\
\vspace{.2in}
 Instituto de F\'{\i}sica Te\'orica, UNESP - S\~ao Paulo State University,\\
 Caixa Postal 70532-2, 01156-970 S\~ao Paulo, SP, Brazil. \\
 05/11/2010\\

\subsection*{\\Abstract}
\end{center}

\baselineskip .1in

The full Dicke model describes a system of $N$ identical two level-atoms coupled to a 
single-mode quantized bosonic field. The model considers rotating and counter-rotating
coupling terms between the atoms and the bosonic field, with coupling constants $g_1$ and $g_2$,
for each one of the coupling terms, respectively. We study finite temperature properties of the model 
using the path integral approach and functional methods. In the thermodynamic limit, 
$N\rightarrow\infty$, the system exhibits phase transition from normal to superradiant phase, 
at some critical values of temperature and coupling constants. We distinguish between three particular cases,
the first one corresponds to the case of rotating wave approximation, 
which $g_1\neq 0$ and $g_2=0$, the second one corresponds to the case of $g_1=0$ and $g_2\neq 0$,
in these two cases the model has a continuous symmetry. The last one, corresponds to the case of 
$g_1\neq 0$ and $g_2\neq 0$, which the model has a discrete symmetry. The phase transition in each case is 
related to the spontaneous breaking of its respective symmetry.
For each one of these three particular cases, we find the asymptotic behaviour of 
the partition function in the thermodynamic limit, and the collective spectrum of the system in the 
normal and the superradiat phase. For the case of rotating wave approximation, and also the case of 
$g_1=0$ and $g_2\neq 0$, in the superradiant phase, the collective spectrum has a zero energy
value, corresponding to the Goldstone mode associated to the continuous symmetry breaking of the model. Our
analyse and results are valid in the limit of zero temperature, $\beta\rightarrow\infty$, in which,
the model exhibits a quantum phase transition.

\vspace{0,1in}
PACS numbers: 03.65.Db, 05.30.Jp, 73.43.Nq, 73.43.Lp

\footnotetext[1]{e-mail: \,aparicio@ift.unesp.br}
\footnotetext[2]{e-mail:\,\,pimentel@ift.unesp.br}

\end{titlepage}

\newpage\baselineskip .18in

\section{Introduction}

\quad $\,\,$
The Dicke model is an interesting spin-boson model because, being
a simple model, exhibits the superradiance effect \cite{dicke}. This model describes a system of $N$ 
identical two level-atoms coupled to a single-mode radiation field, simplified according to the rotating wave
approximation. In this context, the super-radiance is characterized as the coherent spontaneous 
radiation emission with intensity proportional to $N^2$.  
Thermodynamic properties of the Dicke model were studied in the thermodynamic limit, $N\rightarrow\infty$. 
It is found that, the model exhibits a second order phase transition from normal to superradiant phase
at certain critical temperature and sufficiently larger value of the coupling constant 
between the atoms and the field \cite{hepp} \cite{wang}. The influence of the counter-rotating 
term on the thermodynamics of the Dicke model also was studied in the literature, \cite{hepp2} 
\cite{duncan}. Using different coupling constants between
the rotating and the counter-rotating coupling, it is calculated the
critical temperature and the free energy of the model
\cite{hioe} \cite{pimentel1}. We call this generalization of full Dicke model.
Path integral approach and functional methods were used for study spin-boson problems,
finding critical temperature, free energy and collective spectrum of the models, 
in the thermodynamic limit, \cite{moshchi} \cite{yarunin}. With this approach, 
Popov and Fedotov \cite{popov1} \cite{popov2}, rigorously calculated the partition 
function and collective spectrum for the Dicke model in the normal and superradiant phase. 
Relation between the phase transition and continuous symmetry breaking in the Dicke model was
pointed out in reference \cite{popov3}. The full Dicke model was studied using the path integral
approach \cite{aparicio1}, 
here the authors find the asymptotic behaviour of the partition function and collective spectrum
in the normal phase. Using the same approach, thermodynamic properties of some other spin-boson 
models were also studied \cite{aparicio2} \cite{aparicio3}.

In this paper, using the path integral approach and functional methods, we find the 
asymptotic behaviour of the partition function and collective spectrum  
of the full Dicke model in the thermodynamic limit, $N\rightarrow\infty$, in the normal
and super-radiant phase.
The full Dicke model exhibits phase transition from normal to superradiant phase, 
at some critical values of temperature and coupling constants. In our study we 
distinguish three particular cases.
The first one corresponds to the case of rotating wave approximation, 
$g_1\neq 0$ and $g_2=0$, in this case the model has a continuous symmetry, which is associated
to the conservation of the sum of the number excitation of the $N$ atoms with the number 
excitation of the boson field. The second case corresponds to the model with $g_1=0$ and $g_2\neq 0$,
in this case the model also has a continuous symmetry, which is associated to the conservation 
of the difference between the number excitation of the $N$ atoms and the number 
excitation of the boson field. The last one corresponds to the case of 
$g_1\neq 0$ and $g_2\neq 0$, which the model has a discrete symmetry. The phase transition in each case is 
related to the spontaneous breaking of their respective symmetry. For the case of rotating 
wave approximation, and also for the case of $g_1=0$ and $g_2\neq 0$, in the superradiant phase,
the collective spectrum has a zero energy value, corresponding to the Goldstone mode associated to 
the breaking of their respective continuous symmetry. The collective spectrum obtained in this paper 
is valid for the zero temperature limit, corresponding to the case of quantum phase transition.

Practical realization of the full Dicke model in the laboratory was discussed by Dimer {\it et al.} \cite{dimer}.
Since the radiation frequency and energy separation between the two levels of the atoms exceed
the coupling constant strength by many orders of magnitude the counter-rotating terms have a little 
effect on the dynamics. These authors proposed that in cavities with the $N$ qubits, only
one mode of quantized field and classical fields (lasers), it is possible to obtain an effective
Hamiltonian equal to the full Dicke Hamiltonian. It is possible to control the parameters 
in this effective Hamiltonian, and it is possible to operate in the phase transition regime. Other authors stressed the
importance for quantum information technology of experimental realization of generalizations of the 
Dicke model in cavity quantum electrodynamics \cite{harkonen} \cite{baumann}. 

Quantum phase transition of the Dicke model, in the thermodynamic limit, is studied by diagonalizing the
Hamiltonian \cite{hillery}. For this purpose it is applied the Holstein-Primakoff map, which represents the total
angular momentum of the $N$ atoms by a single bosonic field. These author find the collective spectrum
in the normal phase. Similar method was used by Emary and Brandes to study the connection between the 
quantum phase transition and the quantum chaos in the Dicke model without using the rotating wave 
approximation \cite{emary2}. They find the collective spectrum of the model in the normal and superradiant phase,
as another quatities properly of quantum chaos. The relationship between entanglement and quantum phase transition
in the Dicke model was also studied \cite{ent1} \cite{ent2}, the authors find that the atom-field entanglement entropy 
diverges at the critical point of the phase transition. Studies of this relationship between entanglement and quantum 
phase transition for others collective models exist in the literature \cite{vidal}.

This paper is organized as follows. In section 2, we introduce the full Dicke Hamiltonian and study
its symmetries. In section 3, we introduce a map between the spin momentum operators of each atom, with bilinear
forms of fermionic operators, defining the fermion full Dicke model. In section 4, we are able to introduce
the path integral approach for the full Dicke model, using functional methods we obtain
the critical temperature and the asymptotic behaviour of the partition function in some particular cases of the model. 
In section 5, partition function and collective spectrum of the model are presented in the normal phase. In section 6, 
partition function and collective spectrum of the model are presented in the superradiant phase. In 
section 7 we discuss our conclusions. In the paper we use $k_{B}=c=\hbar=1$.

\section{The full Dicke Hamiltonian and symmetries}
\quad $\,\,$ 
The full Dicke model describes a system of $N$ identical two level-atoms coupled to a 
single-mode quantized bosonic field. The model considers rotating and counter-rotating
coupling terms between the atoms and the bosonic field in the Hamiltonian, with coupling constants $g_1$ and $g_2$,
for each one of the coupling terms, respectively. Consequently, the Hamiltonian of the full Dicke model can be 
written as
\begin{eqnarray}
H\,=\,\frac{\Omega}{2}\,\sum_{j=1}^{N}\,\sigma_{(j)}^z+\omega_{0}\,b^{\dagger}\,b\,+\frac{g_1}{\sqrt{N}}
\sum_{j=1}^{N}\, \Bigl(b\,
\sigma_{(j)}^{+}+b^{\dagger}\sigma_{(j)}^{-}\Bigr)+\,\frac{g_2}{\sqrt{N}}
\sum_{j=1}^{N}\, \Bigl(b\,
\sigma_{(j)}^{-}+b^{\dagger}\sigma_{(j)}^{+}\Bigr)\,. 
\label{fullDHamil}
\end{eqnarray}
In above equation we define the operators $\sigma_{(j)}^{\pm}=\frac{1}{2}\,(\sigma_{(j)}^{1} \pm i\,\sigma_{(j)}^{2})$, 
which the operators $\sigma_{(j)}^1$,
$\sigma_{(j)}^2$ and $\sigma_{(j)}^z=\sigma_{(j)}^3$ satisfy the commutation relations $[\sigma_{(j)}^p,\sigma_{(j)}^q]=
2\,\epsilon^{pqr}\,\sigma_{(j)}^r$ with $p,q,r=1,2,3$. Therefore, $[\sigma_{(j)}^+,\sigma_{(j)}^-]=\sigma_{(j)}^z$ and $[\sigma_{(j)}^z,\sigma_{(j)}^{\pm}]=
\pm\,2\,\sigma_{(j)}^{\pm}$. The
$b$ and $b^{\dagger}$ are the boson annihilation and creation
operators of mode excitations that satisfy the usual commutation
relation rules.

Let us define three different operators. The first one, the operator $N$, which is defined by
\begin{eqnarray}
N=b^{\dagger}b+\frac{1}{2}\sum_{i=1}^N\sigma_{(i)}^z\,.
\label{Nexc1}
\end{eqnarray}
The second one, the operator $N_-$, defined by
\begin{eqnarray}
N_-=b^{\dagger}b-\frac{1}{2}\sum_{i=1}^N\sigma_{(i)}^z\,.
\label{N-1}
\end{eqnarray}
Finally, we define the parity operator $\Pi$ by
\begin{eqnarray}
\Pi=e^{i\,\pi\,N}\,, 
\label{parity1}
\end{eqnarray}
with operator $N$ defined in Eq. (\ref{Nexc1}). In particular case of $g_1\neq 0$ and $g_2=0$, 
which corresponds to the rotating wave approximation case, it is possible to show that $[H,N]=0$. 
In particular case of $g_1=0$ and $g_2\neq 0$, it is possible to show that $[H,N_-]=0$.
And it is possible to show that $[H,\Pi]=0$ for arbitrary non-negative values of $g_1$ and 
$g_2$. These commutation relations of the Hamiltonian with each operator defined above, correspond 
to symmetries of the model for each case. It is interesting
to see that, for the case of $g_1\neq 0$ and $g_2\neq 0$, we only have that $[H,\Pi]=0$, it means that 
the system only has parity symmetry. The operators defined by 
$J^p=\frac{1}{2}\sum_{i=1}^N\sigma_{(i)}^p$ with $p=1,2,3$, satisfy the usual angular momentum 
commutation relations. The Hilbert space corresponding to the atoms states can be generated
by the basis $\{|j\,m\rangle\}$ with $j=N/2$ and $m=-j,-j+1,...,j-1,j$; each basis state satisfies 
$J^3|j\,m\rangle =m|j\,m\rangle$ and ${\bf J}^2|j\,m\rangle =j(j+1)|j\,m\rangle$. The Hilbert space, 
which the photon states are defined, can be generated by the basis $\{|n\rangle\}$, with their elements satisfying
$b^{\dagger}b|n\rangle =n|n\rangle$, in this case, $n$ is the number of 
photons. Now we are able to construct a basis for the total system as a tensor product of the above basis
introduced, i.e., the set $\{|n\rangle \otimes |j\,m\rangle\}$. The symmetries mentioned above,
are related with conserved quatities. In the case of $g_1\neq 0$ and $g_2=0$, with $[H,N]=0$, the 
excitation number of the system, $n+m$, is conserved.
It means that the temporal evolution of a state given by $|n\rangle \otimes |j\,m\rangle$ only evolves
toward another states $|n'\rangle \otimes |j\,m'\rangle$ which $n'+m'=n+m$. In similar fashion, for the case 
of $g_1=0$ and $g_2\neq 0$, with $[H,N_-]=0$, the difference of excitation numbers, $n-m$, is conserved. 
When $g_1\neq 0$ and $g_2\neq 0$, which $[H,\Pi]=0$, the value $e^{i\,\pi\, (n+m)}$ is conserved. It means that 
the temporal evolution of a state given by $|n\rangle \otimes |j\,m\rangle$ only evolves toward another states 
$|n'\rangle \otimes |j\,m'\rangle$ with both, $n+m$ and $n'+m'$ being even or $n+m$ and $n'+m'$ being odd.
In all mentioned cases, the phase transition is related to the spontaneous breaking of their respective 
symmetries. In further analysis we shall see that, the symmetry  associated to the commutation relation
$[H,\Pi]=0$ is discrete, and the symmetries associated to the commutation relations $[H,N]=0$
and $[H,N_-]=0$ are continuous symmetries. In cases of continuous symmetry breaking the Goldstone theorem is valid, 
with the appearing of zero energy value in the phase with the symmetry broken.

\section{The fermion full Dicke model}
\quad $\,\,$ 

Let us define the fermion full Dicke model. For this purpose, let us define the raising
and lowering Fermi operators $\alpha^{\dagger}_{i}$, $\alpha_{i}$,
$\beta^{\dagger}_{i}$ and $\beta_{i}$, that satisfy the
anti-commutator relations
$\alpha_{i}\alpha^{\dagger}_{j}+\alpha^{\dagger}_{j}\alpha_{i}
=\delta_{ij}$ and
$\beta_{i}\beta^{\dagger}_{j}+\beta^{\dagger}_{j}\beta_{i}
=\delta_{ij}$. In this analysis, we use a representation of
the operators $\sigma_{(j)}^z$, $\sigma_{(j)}^+$ and $\sigma_{(j)}^-$ by the following bilinear
combination of Fermi operators, $\alpha^{\dagger}_{i}\alpha_{i}
-\beta^{\dagger}_{i}\beta_{i}$, $\alpha^{\dagger}_{i}\beta_{i}$
and $\beta^{\dagger}_{i}\alpha_{i}$, the correspondence is given by
\begin{equation}
\sigma_{(i)}^{z}\longrightarrow \alpha_{i}^{\dagger}\alpha_{i}
-\beta_{i}^{\dagger}\beta_{i}\, , \label{34}
\end{equation}
\begin{equation}
\sigma_{(i)}^{+}\longrightarrow \alpha_{i}^{\dagger}\beta_{i}\, ,
\label{35}
\end{equation}
and
\begin{equation}
\sigma_{(i)}^{-}\longrightarrow \beta_{i}^{\dagger}\alpha_{i}\, .
\label{36}
\end{equation}
Using this representation given in Eq. (\ref{34}), Eq. (\ref{35}) and Eq. (\ref{36})
in the full Dicke Hamiltonian given by Eq. (\ref{fullDHamil}), we define 
the Hamiltonian of the fermion full Dicke model $H_F$. So that, we have
\begin{eqnarray}
H_F=\omega_0\; b^{\dagger}b+ \frac{\Omega}{2}\sum_{i=1}^N
\Bigl(\alpha_i^{\dagger}\alpha_i -\beta_i^{\dagger}\beta_i\Bigr)
+\frac{g_1}{\sqrt{N}}\sum_{i=1}^N
\Bigl(b\,\alpha_i^{\dagger}\beta_i \,+\, b^{\dagger}\,
\beta_i^{\dagger}\alpha_i \Bigr)+\,
\frac{g_2}{\sqrt{N}}\sum_{i=1}^N
\Bigl(b^{\dagger}\,\alpha_i^{\dagger}\beta_i \,+\, b\,
\beta_i^{\dagger}\alpha_i \Bigr)\, . \label{37}
\end{eqnarray}

We are interested in studying thermodynamic properties of the system, therefore we must find the partition
function $Z$. It is important to note that Hamiltonians $H$ and $H_F$ are defined in different spaces. 
Each operator $\sigma_i^{\alpha}$ appearing in the Hamiltonian $H$ acts on two-dimensional Hilbert space, 
notwithstanding, Fermi operators $\alpha^{\dagger}_{i}$, $\alpha_{i}$,
$\beta^{\dagger}_{i}$ and $\beta_{i}$, appearing in the Hamiltonian
$H_F$ act on four-dimensional Fock space. The following property relates the 
partition function of the full Dicke model with the partition function of the fermion full Dicke model:
\begin{eqnarray}
Z=Tr\Bigl(\exp(-\beta\,H)\Bigr)=i^N\,Tr\left(\exp\left(-\beta\,H_F-\frac{i\pi}{2}\,N_F\right)\right)\,.
\label{partitionsfunctions}
\end{eqnarray}
In this last relation $H$ is given by the Eq. (\ref{fullDHamil}), $H_F$ is given by Eq. (\ref{37}) 
and the operator $N_F$ is defined by
\begin{eqnarray}
N_F=\sum_{i=1}^{N}(\alpha_{i}^{\dagger}\alpha_{i}+\beta_{i}^{\dagger}\beta_{i})\,.
\end{eqnarray}
The traces used in Eq. (\ref{partitionsfunctions}) for each Hamiltonians are carried over their repective 
spaces. The relation given by Eq. (\ref{partitionsfunctions}) let us express the partition function of the
full Dicke model $Z$ using the fermion full Dicke Hamiltonian given by Eq. (\ref{37}).

\section{The partition function with path integral approach}

In this section we perform calculations in order to obtain an asymptotic expression for the partition function $Z$ 
of the full Dicke model in the limit of $N\rightarrow\infty$. For this purpose we use path integral approach
and functional methods. Let us define the Euclidean action $S$ of the full Dicke model in the following form
\begin{equation}
S=\int_0^{\beta} d\tau \left(b^*(\tau)\,\partial_{\tau}b(\tau)+ \sum_{i=1}^{N}
\Bigl(\alpha^*_i(\tau)\,\partial_{\tau}\alpha_i(\tau)
+\beta^*_i (\tau)\,\partial_{\tau}\beta_i(\tau)\Bigr)\right) -\int_0^{\beta}d\tau H_{F}(\tau)\,,
\label{66}
\end{equation}
the Hamiltonian $H_{F}$ is the full Hamiltonian for the full fermion
Dicke model, which is given by
\begin{eqnarray}
H_{F}(\tau)\,=\,\omega_{0}\,b^{\,*}(\tau)\,b(\tau)\,+
\,\frac{\Omega}{2}\,\displaystyle\sum_{i\,=\,1}^{N}\,
\biggl(\alpha^{\,*}_{\,i}(\tau)\,\alpha_{\,i}(\tau)\,-
\,\beta^{\,*}_{\,i}(\tau)\beta_{\,i}(\tau)\biggr)\,+
\nonumber\\
+\,\frac{g_{\,1}}{\sqrt{N}}\,\displaystyle\sum_{i\,=\,1}^{N}\,
\biggl(\alpha^{\,*}_{\,i}(\tau)\,\beta_{\,i}(\tau)\,b(\tau)\,+
\alpha_{\,i}(\tau)\,\beta^{\,*}_{\,i}(\tau)\,b^{\,*}(\tau)\,\biggr)\,+
\nonumber\\
+\,\frac{g_{\,2}}{\sqrt{N}}\,\displaystyle\sum_{i\,=\,1}^{N}\,
\biggl(\alpha_{\,i}(\tau)\,\beta^{\,*}_{\,i}(\tau)\,b(\tau)\,+
\,\alpha^{\,*}_{\,i}(\tau)\,\beta_{\,i}(\tau)\,b^{\,*}(\tau)\biggr).
\label{66a}
\end{eqnarray}
Let us define the formal quotient of the partition function of the full Dicke
model and the partition function of the free Dicke model.
Therefore we are interested in calculating the following quantity
\begin{equation}
\frac{Z}{Z_0}=\frac{\int [d\eta]\,\exp{\left(\,S-\frac{i\pi}{2\beta}\int_0^{\beta}n(\tau)d\tau\right)}}{\int
[d\eta]\,\exp{\left(\,S_{0}-\frac{i\pi}{2\beta}\int_0^{\beta}n(\tau)d\tau\right)}}\, , 
\label{65}
\end{equation}
the function $n(\tau)$ is defined by
\begin{eqnarray}
n(\tau)=\sum_{i=1}^{N}
\Bigl(\alpha^{\,*}_i(\tau)\,\alpha_i(\tau)+\beta^{\,*}_i(\tau)\beta_i(\tau)\Bigr)\,,
\end{eqnarray} 
$S=S(b,b^*,\alpha,\alpha^{\dagger},\beta,\beta^{\dagger})$
is the Euclidean action of the full Dicke model
given by Eq. (\ref{66}),
$S_0=S_{0}(b,b^*,\alpha,\alpha^{\dagger},\beta,\beta^{\dagger})$
is the free Euclidean action for the free single bosonic mode and
the free atoms, i.e., the expression of the complete action $S$
taking $g_1=g_2=0$ and finally $[d\eta]$ is the functional
measure.
The functional integrals involved in Eq. (\ref{65}), are
functional integrals with respect to the complex functions
$b^*(\tau)$ and $b(\tau)$ and Fermi fields
$\alpha_i^*(\tau)$, $\alpha_i(\tau)$, $\beta_i^*(\tau)$ and
$\beta_i(\tau)$. Since we are using thermal equilibrium boundary
conditions, in the imaginary time formalism, the integration
variables in Eq. (\ref{65}) obey periodic boundary conditions for
the Bose field, i.e., $b(\beta)=b(0)$ and anti-periodic boundary
conditions for Fermi fields i.e., $\alpha_i(\beta)=-\alpha_i(0)$
and $ \beta_i(\beta)=-\beta_i(0)$.

In section 2, we have analysed the symmetry of the model studying commutation 
relations between the Hamiltonian given by Eq. (\ref{fullDHamil}) with 
some operators defining the symmetry.
Now we are able to analise the symmetry of the model studying the invariance of the action
given by Eq. (\ref{66}) under symmetry transformations. In this way, let us introduce
the following field transformation
\begin{eqnarray}
\begin{array}{ccc}
b(\tau)\rightarrow\,e^{i\,\gamma}\,b(\tau)\,,\;\; & \alpha(\tau)\rightarrow\,e^{i\,\theta}\,\alpha(\tau)\,,\;\; &
\beta(\tau)\rightarrow\,e^{i\,\phi}\,\beta(\tau)\,,\\
\;\;\;\;b^*(\tau)\rightarrow\,e^{-i\,\gamma}\,b^*(\tau)\,,\;\; & \;\;\;\;\alpha^*(\tau)\rightarrow\,e^{-i\,\theta}\,\alpha^*(\tau)\,,\;\; &
\;\;\;\;\beta^*(\tau)\rightarrow\,e^{-i\,\phi}\,\beta^*(\tau)\,.
\label{symmtranf1}
\end{array}
\end{eqnarray}
In the case of $g_1\neq 0$ and $g_2=0$, corresponding to the case of rotating wave approximation, its respective
action is invariant under tranformation given by Eq. (\ref{symmtranf1}), taking $\gamma=\theta-\phi$.
In the case of $g_1=0$ and $g_2\neq 0$, its corresponding action is invariant under tranformation given by 
Eq. (\ref{symmtranf1}), taking $\gamma=\phi-\theta$. Finally, in the case of $g_1\neq 0$ and $g_2\neq 0$, its 
corresponding action is invariant under tranformation given by Eq. (\ref{symmtranf1}), with $\gamma=\theta-\phi=0$ or
$\gamma=\theta-\phi=\pi$. In the two first cases, the case of $g_1\neq 0$ and $g_2=0$, and 
the case of $g_1=0$ and $g_2\neq 0$, their respective actions are invariant under 
continuous transformation, $U(1)$, of the boson field $b(\tau)$. In the case of $g_1\neq 0$ and $g_2\neq 0$, 
its action is invariant under discrete transformations, $Z_2$, of the boson field $b(\tau)$, i. e., 
$b(\tau)\rightarrow b(\tau)$ and $b(\tau)\rightarrow -b(\tau)$.

Following with the purpose of calculating the quantity $\frac{Z}{Z_0}$ given by Eq. (\ref{65}), 
let us use the following transformation
\begin{eqnarray}
\begin{array}{cc}
\alpha_i(\tau)\rightarrow e^{\frac{i\pi}{2\beta}t}\,\alpha_i(\tau)\,,\;\;\;  &  
\alpha_i^*(\tau)\rightarrow e^{-\,\frac{i\pi}{2\beta}t}\,\alpha_i^*(\tau)\,,\\
\beta_i(\tau)\rightarrow e^{\frac{i\pi}{2\beta}t}\,\beta_i(\tau)\,,\;\;\;  &  
\beta_i^*(\tau)\rightarrow e^{-\,\frac{i\pi}{2\beta}t}\,\beta_i^*(\tau)\,.
\end{array}
\label{trans2}
\end{eqnarray}
With this last transformation, the term $n(\tau)$ appearing in Eq. (\ref{65}) can be dropped. Therefore, applying
the tranformation given by Eq. (\ref{trans2}) into the expression given by Eq. (\ref{65}), we obtain that
\begin{equation}
\frac{Z}{Z_0}=\frac{\int [d\eta]\,e^S}{\int[d\eta]\,e^{S_0}}\,. 
\label{65n}
\end{equation}
In Eq. (\ref{65n}), the Bose field obeys periodic boundary conditions, i.e., $b(\beta)=b(0)$,
and the Fermi fields obey the following boundary conditions: 
\begin{eqnarray}
\begin{array}{cc}
\alpha_i(\beta)=i\,\alpha_i(0)\,,\;\;\;& 
\alpha_i^*(\beta)=-\,i\,\alpha_i^*(0)\,,\\
\beta_i(\beta)=i\,\beta_i(0)\,,\;\;\;&
\beta_i^*(\beta)=-\,i\,\beta_i^*(0)\,.
\end{array}
\label{nbound}
\end{eqnarray}

The free action for the single mode bosonic field $S_{B0}(b)$ is
given by
\begin{equation}
S_{B0}(b) = \int_{0}^{\beta} d\tau\; b^{*}(\tau)\,\Bigl(
\partial_{\tau}-\omega_{0}\Bigr)\,b(\tau)\, . \label{67}
\end{equation}
Then we can write the action $S$ of the full fermion Dicke
model, given by Eq. (\ref{66}), using the free action for the
single mode bosonic field $S_{B0}(b)$ defined by Eq. (\ref{67}), plus
an additional term that can be expressed in matrix form.
Therefore the total action $S$ can be written as
\begin{equation}
S = S_{B0}(b) +  \int_{0}^{\beta} d\tau\,\sum_{i=1}^{N}\,
\rho^{\dagger}_{i}(\tau)\, M(b^{*},b)\,\rho_{i}(\tau)\, ,
\label{68}
\end{equation}
the column matrix $\rho_{\,i}(\tau)$ is given in terms of
Fermi field operators in the following way
\begin{eqnarray}
\rho_{\,i}(\tau) &=& \left(
\begin{array}{c}
\beta_{\,i}(\tau) \\
\alpha_{\,i}(\tau)
\end{array}
\right),
\nonumber\\
\rho^{\dagger}_{\,i}(\tau) &=& \left(
\begin{array}{cc}
\beta^{*}_{\,i}(\tau) & \alpha^{*}_{\,i}(\tau)
\end{array}
\right) \label{69a}
\end{eqnarray}
and the matrix $M(b^{*},b)$ is given by
\begin{equation}
M(b^{*},b) = \left( \begin{array}{cc}
L & (N)^{-1/2}\,\biggl(g_{1}\,b^{*}\,(\tau) + g_{2}\,b\,(\tau)\biggr)\\
(N)^{-1/2}\,\biggl(g_{1}\,b\,(\tau) + g_{2}\,b^{*}\,(\tau)\biggr)
&
L_*
\end{array} \right)\,,
\label{69b}
\end{equation}
the operators $L$ and $L_*$ are defined by $\partial_{\tau} + \Omega/2$ and 
$\partial_{\tau} - \Omega/2$ respectively.
Substituting the action $S$ given by Eq. (\ref{68}) in the functional integral form of the partition 
function given by Eq. (\ref{65n}) we see that this functional integral is Gaussian in the Fermi fields.
Now, let us begin integrating with respect these Fermi fields, therefore we obtain
\begin{eqnarray}
Z=\int[d\eta(b)]\,e^{S_{B0}} \Bigl(\det{M(b^{*},b)}\Bigr)^N\,,
\label{Zop}
\end{eqnarray}
in this case, $[d\eta(b)]$ is the functional measure only for the bosonic field.
With the help of the following property for matrices with operator components
\begin{eqnarray}
\det\left(\begin{array}{cc}
A&B\\
C&D
\end{array}
\right)=\det\left(AD-ACA^{-1}B\right)\,,
\label{mazprop}
\end{eqnarray}
and determinant properties, we have that
\begin{eqnarray}
\det{M(b^{*},b)}=\det{\Bigl(LL_*\Bigr)}\,\det{\left(1-N^{-1}L_*^{-1}
\Bigl(g_1\,b+ g_2\,b^*\Bigr)L^{-1}\Bigl(g_1\,b^*+ g_2\,b\Bigr)\right)}\,.
\label{Mop1}
\end{eqnarray}
Substituting Eq. (\ref{Zop}) and Eq. (\ref{Mop1}) in Eq. (\ref{65n}), we have that
\begin{eqnarray}
\frac{Z}{Z_0}=\frac{Z_A}{\int[d\eta(b)]\,e^{S_{B0}}}\,,
\label{ZA0}
\end{eqnarray}
with $Z_A$ defined by
\begin{eqnarray}
Z_A=\int[d\eta(b)]\exp{\left(S_{B0}+N\,tr\ln\biggl(1-N^{-1}L_*^{-1}
\Bigl(g_1\,b+ g_2\,b^*\Bigr)L^{-1}\Bigl(g_1\,b^*+ g_2\,b\Bigr)\biggr)\right)}\,.
\label{ZA1}
\end{eqnarray}

We are interested in knowing the asymptotic behaviour of the quotient $\frac{Z}{Z_0}$ in the
thermodynamic limit, i. e., $N\rightarrow\infty$. With this intention, we analyse 
the asymptotic behaviour of the last defined expression $Z_A$. First, let us scale the bosonic
field by $b\rightarrow\sqrt{N}\,b$ and $b^*\rightarrow\sqrt{N}\,b^*$, so that we get
\begin{eqnarray}
Z_A=A(N)\int[d\eta(b)]\exp{\left(N\,\Phi(b^*,b)\right)}\,,
\label{ZA2}
\end{eqnarray}
with the function $\Phi(b^*,b)$ defined by
\begin{eqnarray}
\Phi(b^*,b)=S_{B0}+tr\ln\biggl(1-L_*^{-1}
\Bigl(g_1\,b+ g_2\,b^*\Bigr)L^{-1}\Bigl(g_1\,b^*+ g_2\,b\Bigr)\biggr)\,.
\label{fi1}
\end{eqnarray}
The term $A(N)$ in Eq. (\ref{ZA2}) comes from transforming the functional measure $[d\eta(b)]$ under scaling  
the bosonic field by $b\rightarrow\sqrt{N}\,b$ and $b^*\rightarrow\sqrt{N}\,b^*$. The asymptotic behaviour 
of the integral functional appearing in Eq. (\ref{ZA2})
when $N\rightarrow\infty$, can be obtained by using the method of steepest descent \cite{amit}. In this method, 
we expand the function $\Phi(b^*,b)$ around the point $b(\tau)=b_0(\tau)$ and 
$b^*(\tau)=b^*_0(\tau)$, which can be of two kinds. One kind that makes $Re(\Phi(b^*,b))$ maximum, 
and the other kind is defined as saddle point.
We consider the first terms of the expansion in the integral functional, which are the leading 
terms for the value of the integral function. We can find the maximum points, or
saddle points, finding the stationary points. The stationary points are solution of the following equations $\frac{\delta\,
\Phi(b^*,b)}{\delta\,b(\tau)}=0$ and $\frac{\delta\,\Phi(b^*,b)}{\delta\,b^*(\tau)}=0$.
For the full Dicke model, the stationary points are constant functions $b(\tau)=b_0$ and $b^*(\tau)=b^*_0$.
It is not difficult to show that for $\beta\leq\beta_c$ the stationary point
is given by $b_0=b_0^*=0$, which is a maximum point. The critical value $\beta_c$ is obtained by solving the following equation
\begin{eqnarray}
\frac{\omega_0\,\Omega}{(g_1+g_2)^2}=\tanh\left(\frac{\beta_c\,\Omega}{2}
\right)\,.
\label{tcrit}
\end{eqnarray}
In this last equation, it is possible to find some solution for $\beta_c$, in the
case of $(g_1+g_2)^2>\omega_0\Omega$. With this condition the system undergoes a phase transition.
When the system has $\beta<\beta_c$ we say that the system is in the normal phase.
For $\beta>\beta_c$ the stationary points $b(\tau)=b_0$ and $b^*(\tau)=b^*_0$ satisfy the following 
equation
\begin{eqnarray}
\frac{\omega_0\,\Omega_{\Delta}}{(g_1+g_2)^2}=\tanh\left(\frac{\beta\,\Omega_{\Delta}}{2}\right)\,,
\label{tra1}
\end{eqnarray}
with $\Omega_{\Delta}$ defined by
\begin{eqnarray}
\Omega_{\Delta}=
\sqrt{\Omega^2+4\,(g_1+g_2)^2\,|b_0|^2}\,.
\label{omegadelta}
\end{eqnarray}
Phase transition happens if it is possible to find some real solution for $|b_0|\neq 0$ in Eq. (\ref{tra1}).
It is only possible when $(g_1+g_2)^2>\omega_0\,\Omega$ and $\beta>\beta_c$. In the case of $g_1\neq 0$ and 
$g_2=0$, and also in the case of $g_1=0$ and $g_2\neq 0$, the maximum points are a continuous set of values given
by the expression $b_0=\rho\,e^{i\,\phi}$ and $b^*_0=\rho\,e^{-i\,\phi}$ with $\phi\in [0,2\pi)$ and 
$\rho=|b_0|$, with $|b_0|$ defined by Eq. (\ref{tra1}). In the case of $g_1\neq 0$ and $g_2\neq 0$, we have 
two maximum points, which are given by $b^*_0=b_0=\pm |b_0|$, with $|b_0|$ defined by 
Eq. (\ref{tra1}). When the system has $\beta>\beta_c$ we say that the system is in the superradiant phase. 

Let us continue, with the computation of the asymptotic behaviour for the integral functional appearing in Eq. (\ref{ZA2}),
for the thermodynamic limit, $N\rightarrow\infty$. In following steps, we shall find this asymptotic behaviour 
when we only have one maximum point defined by $b_0=b^*_0$. The resulting expressions will be useful for the normal phase
of the full Dicke model, and also for the superradiant phase in the case of $g_1\neq 0$ and $g_2\neq 0$. We consider 
the two first leading terms in the integral function appearing in Eq. (\ref{ZA2}) coming from the expansion
of $\Phi(b^*,b)$ around the maximal value $b^*_0=b_0$, this expansion is given by
\begin{eqnarray}
\Phi(b^*,b)=\Phi(b^*_0,b_0)+\frac{1}{2}\int_0^{\beta}d\tau_1\,d\tau_2\,
(b^*(\tau_1)-b^*_0\,,\,b(\tau_1)-b_0)\,M_{\Phi}
\left(\begin{array}{c}
b^*(\tau_2)-b^*_0\\
b(\tau_2)-b_0
\end{array}\right)\,,
\label{fi2}
\end{eqnarray}
the matriz $M_{\Phi}$, is given by
\begin{eqnarray}
M_{\Phi}=\left(\begin{array}{cc}
\frac{\delta^2\Phi(b^*,b)}{\delta b^*(\tau_1)\,\delta b^*(\tau_2)}&\frac{\delta^2\Phi(b^*,b)}{\delta b^*(\tau_1)\,\delta b(\tau_2)}\\
\frac{\delta^2\Phi(b^*,b)}{\delta b(\tau_1)\,\delta b^*(\tau_2)}&\frac{\delta^2\Phi(b^*,b)}{\delta b(\tau_1)\,\delta b(\tau_2)}
\end{array}\right)
\Biggr|_{b^*=b=b_0}\,.
\label{Mfi}
\end{eqnarray}
Substituting this expansion given by Eq. (\ref{fi2}) in Eq. (\ref{ZA2}) we obtain
\begin{eqnarray}
Z_A=e^{N\Phi(b^*_0,b_0)}\int[d\eta(b)]\exp{\left(\frac{1}{2}\int_0^{\beta}d\tau_1\,d\tau_2\,
\Bigl(b^*(\tau_1)\,,\,b(\tau_1)\Bigr)\,M_{\Phi}
\left(\begin{array}{c}
b^*(\tau_2)\\
b(\tau_2)
\end{array}\right)\right)}\,,
\label{ZA3}
\end{eqnarray}
to obtain the last expression, we have applied the transformation $b(\tau)\rightarrow \Bigr(b(\tau)+b_0\Bigl)/\sqrt{N}$ and 
$b^*(\tau)\rightarrow \Bigr(b^*(\tau)+b^*_0\Bigl)/\sqrt{N}$ in the functional integral involved. In order to make easier 
the integration of the functional integral given by Eq. (\ref{ZA2}), let us use the following transformation
\begin{eqnarray}
c\,(\tau)&=&\alpha\,\Bigl(g_2\,b(\tau)+g_1\,b^*(\tau)\Bigr)\nonumber\\
c^*(\tau)&=&\alpha\,\Bigl(g_1\,b(\tau)+g_2\,b^*(\tau)\Bigr)\,,
\label{Trv1}
\end{eqnarray}
the parameter $\alpha$ defined by the equation $\alpha^2=(g_2^2-g_1^2)^{-1}$. It is worth mentioning that, the Jacobian of this transformation 
is $1$. Applying this transformation in Eq. (\ref{ZA2}) we obtain that
\begin{eqnarray}
Z_A=A(N)\int[d\eta(c)]\exp{\left(N\,\Phi_I(c^*,c)\right)}\,,
\label{ZA4}
\end{eqnarray}
the function $\Phi_I(c^*,c)$ is given by 
\begin{eqnarray}
\Phi_I(c^*,c)&=&\alpha^2\int_0^{\beta}d\tau\,\Bigl(g_1\,c(\tau)-g_2\,c^*(\tau)\Bigr)\times\nonumber\\
&&\times\Bigl(\partial_{\tau}-\omega_0\Bigr)\,
\Bigl(g_1\,c^*(\tau)- g_2\,c(\tau)\Bigr)+tr\ln\biggl(1-\alpha^{-2}L_*^{-1}c^*L^{-1}c\biggr)\,.
\label{fi3}
\end{eqnarray}
The maximum point corresponds to $c^*_0=c_0=\alpha(g_1+g_2)b_0$, the point $b^*_0=b_0$ corresponds to a maximum for 
the function $Re(\Phi(b^*,b))$. Using the same expansion given in Eq. (\ref{fi2}) 
for $\Phi_I(c^*,c)$ and substituting in Eq. (\ref{ZA4}) we obtain that
\begin{eqnarray}
Z_A=e^{N\Phi(b^*_0,b_0)}\int[d\eta(c)]\exp{\left(\frac{1}{2}\int_0^{\beta}d\tau_1\,d\tau_2\,
\Bigl(c^*(\tau_1)\,,\,c(\tau_1)\Bigr)\,M_{\Phi_I}
\left(\begin{array}{c}
c^*(\tau_2)\\
c(\tau_2)
\end{array}\right)\right)}\,,
\label{ZA5}
\end{eqnarray}
we have used the identity $\Phi_I(c^*_0,c_0)=\Phi(b^*_0,b_0)$, and the matrix $M_{\Phi_I}$ is 
defined by
\begin{eqnarray}
M_{\Phi_I}=\left(\begin{array}{cc}
\frac{\delta^2\Phi_I(c^*,c)}{\delta c^*(\tau_1)\,\delta c^*(\tau_2)}&\frac{\delta^2\Phi_I(c^*,c)}{\delta c^*(\tau_1)\,\delta c(\tau_2)}\\
\frac{\delta^2\Phi_I(c^*,c)}{\delta c(\tau_1)\,\delta c^*(\tau_2)}&\frac{\delta^2\Phi_I(c^*,c)}{\delta c(\tau_1)\,\delta c(\tau_2)}
\end{array}\right)
\Biggr|_{c^*=c=c_0}\,.
\label{MfiI}
\end{eqnarray}

At this level, it is convenient to use Fourier representation of the field $c(\tau)$ in the functional integral 
Eq. (\ref{ZA5}). From boundaries conditions of the bosonic field $b(\tau)$ and from Eq. (\ref{Trv1}),
we deduce that $c(\tau)$ and $c^*(\tau)$ satisfies periodic boundary conditions $c(\beta)=c(0)$ and $c^*(\beta)=c^*(0)$
respectively. Therefore Fourier representation of $c(\tau)$ and $c^*(\tau)$ are given by
\begin{eqnarray}
c(\tau)&=&\frac{1}{\sqrt{\beta}}\sum_{\omega}c(\omega)e^{i\omega\tau}\,,\nonumber\\
c^*(\tau)&=&\frac{1}{\sqrt{\beta}}\sum_{\omega}c^*(\omega)e^{-i\omega\tau}\,,
\label{fourier1}
\end{eqnarray}
the parameter $\omega$ takes the values: $2\pi\,n/\beta$, with $n$ being all the integers. These values
correspond to the Matsubara frenquencies for bosonic fields. Substituting this Fourier representation,
Eq. (\ref{fourier1}), in Eq. (\ref{ZA5}), we obtain that
\begin{eqnarray}
Z_A=e^{N\Phi(b^*_0,b_0)}\int[d\eta(c)]\exp{\left(\frac{1}{2}\sum_{\omega_1\omega_2}
\Bigl(c^*(\omega_1)\,,\,c(\omega_1)\Bigr)\,\delta^2\Phi(\omega_1,\omega_2)
\left(\begin{array}{c}
c^*(\omega_2)\\
c(\omega_2)
\end{array}\right)\right)}\,,
\label{ZA6}
\end{eqnarray}
with $\delta^2\Phi(\omega_1,\omega_2)$ being defined by
\begin{eqnarray}
\delta^2\Phi(\omega_1,\omega_2)=\left(\begin{array}{cc}
\delta^2\Phi_{11}(\omega_1,\omega_2) & \delta^2\Phi_{12}(\omega_1,\omega_2)\\
\delta^2\Phi_{21}(\omega_1,\omega_2) & \delta^2\Phi_{22}(\omega_1,\omega_2)
\end{array}\right)\,,
\label{deltafi}
\end{eqnarray}
and each component of this matrix satisfies
\begin{eqnarray}
\delta^2\Phi_{11}(\omega_1,\omega_2)&=&\frac{1}{\beta}\int_0^{\beta}d\tau_1d\tau_2\,\,
e^{-i\omega_1\tau_1}\,\frac{\delta^2\Phi_I(c^*,c)}{\delta c^*(\tau_1)\,\delta c^*(\tau_2)}
\Biggr|_{c=c^*=c_0}e^{-i\omega_2\tau_2}\,,\nonumber\\
\delta^2\Phi_{12}(\omega_1,\omega_2)&=&\frac{1}{\beta}\int_0^{\beta}d\tau_1d\tau_2\,\,
e^{-i\omega_1\tau_1}\,\frac{\delta^2\Phi_I(c^*,c)}{\delta c^*(\tau_1)\,\delta c(\tau_2)}
\Biggr|_{c=c^*=c_0}e^{i\omega_2\tau_2}\,,\nonumber\\
\delta^2\Phi_{21}(\omega_1,\omega_2)&=&\frac{1}{\beta}\int_0^{\beta}d\tau_1d\tau_2\,\,
e^{i\omega_1\tau_1}\,\frac{\delta^2\Phi_I(c^*,c)}{\delta c(\tau_1)\,\delta c^*(\tau_2)}
\Biggr|_{c=c^*=c_0}e^{-i\omega_2\tau_2}\,,\nonumber\\
\delta^2\Phi_{22}(\omega_1,\omega_2)&=&\frac{1}{\beta}\int_0^{\beta}d\tau_1d\tau_2\,\,
e^{i\omega_1\tau_1}\,\frac{\delta^2\Phi_I(c^*,c)}{\delta c(\tau_1)\,\delta c(\tau_2)}
\Biggr|_{c=c^*=c_0}e^{i\omega_2\tau_2}\,.
\label{deltaficomp}
\end{eqnarray}
In this Fourier representation of the functional integral given by Eq. (\ref{ZA6}), the integral
measure $[d\eta(c)]$ takes the tractable form $\prod_{\omega}{dc^*(\omega)\,dc^*(\omega)}$. 
Using the expression for $\Phi_I(c^*,c)$ given in Eq. (\ref{fi3}), we can calculate the matriz
$\delta^2\Phi(\omega_1,\omega_2)$ with components given by Eq. (\ref{deltaficomp}). Performing
these calculations we obtain that
\begin{eqnarray}
\delta^2\Phi_{11}(\omega_1,\omega_2)&=&\delta^2\Phi_{12}(\omega_1,\omega_2)=
\delta_{\omega_1\,,\,-\omega_2}\,R(\omega_1)\,,\nonumber\\
\delta^2\Phi_{21}(\omega_1,\omega_2)&=&\delta^2\Phi_{22}(\omega_1,\omega_2)=
\delta_{\omega_1\,,\,\omega_2}\,S(\omega_1)\,,
\label{deltaficomp1}
\end{eqnarray}
with $\delta_{\omega_1\,,\,\omega_2}$ being the delta Kronecker and the functions $R(\omega)$ 
and $S(\omega)$ are given by
\begin{eqnarray}
R(\omega)&=&2\,\omega_0\,g_1\,g_2\,\alpha^2-\frac{(\,\Omega^2_{\Delta}-\Omega^2)\,\alpha^{-2}}
{2\,\Omega_{\Delta}(\omega^2+\Omega^2_{\Delta})}\tanh{\left(\frac{\beta\,\Omega_{\Delta}}{2}\right)}\,,\nonumber\\
S(\omega)&=&i\,\omega\left(1-\frac{\Omega\,\alpha^{-2}}
{\Omega_{\Delta}(\omega^2+\Omega^2_{\Delta})}\tanh{\left(\frac{\beta\,\Omega_{\Delta}}{2}\right)}\right)+\nonumber\\
&-&
\omega_0\,(\,g_1^2+g_2^2\,)\,\alpha^2+\frac{(\,\Omega^2_{\Delta}+\Omega^2)\,\alpha^{-2}}
{2\,\Omega_{\Delta}\,(\omega^2+\Omega^2_{\Delta})}\tanh{\left(\frac{\beta\,\Omega_{\Delta}}{2}\right)}\,.
\label{RS}
\end{eqnarray}
The expression for $\Omega_{\Delta}$ is given by Eq. (\ref{omegadelta}).
Substituting the matriz $\delta^2\Phi(\omega_1,\omega_2)$, with components given by Eq. (\ref{deltaficomp1}), in
the functional integral appearing in $Z_A$, given by Eq. (\ref{ZA6}), we obtain that
\begin{eqnarray}
Z_A=e^{N\Phi(b^*_0,b_0)}\int[d\eta(c)]\exp{\sum_{\omega}\left(\,
S(\omega)\,c(\omega)\,c^*(\omega)+\frac{1}{2}\,R(\omega)\,
\Bigl(\,c(\omega)\,c(-\omega)+c^*(\omega)\,c^*(-\omega)\,\Bigr)\right)}\,.
\label{ZA7}
\end{eqnarray}
Performing this Gaussian functional integral, we finally obtain that
\begin{eqnarray}
Z_A=e^{N\Phi(b^*_0,b_0)}\frac{2\,\pi\,i}{(S^2(0)-R^2(0))^{1/2}}\,\,
\prod_{\omega\geq 1}\frac{(\,2\,\pi\,i\,)^2}{\,S(\omega)\,S(-\omega)\,-\,R^2(\omega)}\,.
\label{ZA8}
\end{eqnarray}

In order to find the asymptotic behaviour of $\frac{Z}{Z_0}$ when $N\rightarrow\infty$, we must 
calculate $\int{[d\eta(b)]\,e^{S_{B0}}}$ appearing in Eq. (\ref{ZA0}). Using the free bosonic 
action $S_{B0}$ given by Eq. (\ref{67}), we obtain that
\begin{eqnarray}
\int{[d\eta(b)]\,e^{S_{B0}}}=\prod_{\omega}\,\frac{2\,\pi\,i}{\omega_0-i\,\omega}\,.
\label{sb0}
\end{eqnarray}
Substituting Eq. (\ref{ZA8}) and Eq. (\ref{sb0}) in Eq. (\ref{ZA0}) we have that
\begin{eqnarray}
\frac{Z}{Z_0}=e^{N\Phi(b^*_0,b_0)}\frac{1}{(H(0))^{1/2}}\,\,
\prod_{\omega\geq 1}\,\frac{1}{\,H(\omega)}\,,
\label{Zfin}
\end{eqnarray}
in this last equation, the function $H(\omega)$ is given by
\begin{eqnarray}
H(\omega)=\frac{S(\omega)\,S(-\omega)-R^2(\omega)}{\omega^2+\omega^2_0}\,.
\label{H1}
\end{eqnarray}
The Eq. (\ref{RS}) gives the expresions for the functions $S(\omega)$ and $R(\omega)$, substituting
these functions in Eq. (\ref{H1}) we obtain that
\begin{eqnarray}
&&H(\omega)=\,1\,+\,\frac{(\,g_1^2-g_2^2\,)^2\,\Omega^2}{\Omega_{\Delta}^2\,(\omega^2+\Omega_{\Delta}^2)\,
(\omega^2+\omega^2_0)}\tanh^2\left(\frac{\beta\,\Omega_{\Delta}}{2}\right)\,+\nonumber\\
&+&\frac{2\,(g_1^2-g_2^2\,)\,\Omega\,\omega^2\,-\,(\,g_1^2+g_2^2\,)\,(\Omega^2+\Omega_{\Delta}^2)\,\omega_0\,+\,
2\,g_1\,g_2\,(\Omega_{\Delta}^2-\Omega^2)\,\omega_0}{\Omega_{\Delta}\,(\omega^2+\Omega_{\Delta}^2)\,
(\omega^2+\omega^2_0)}\, \tanh\left(\frac{\beta\,\Omega_{\Delta}}{2}\right)\,.
\label{H2}
\end{eqnarray}
The expression, given by Eq. (\ref{Zfin}), with $H(\omega)$ given by Eq. (\ref{H2}), provides a 
valid expression for the quotient $\frac{Z}{Z_0}$ in the normal phase, and also in the superradiant 
phase for the particular case of $g_1\neq 0$ and $g_2\neq 0$.

\section{Normal phase: $\beta <\beta_c$}
In the normal phase, $\beta <\beta_c$, from Eq. (\ref{tra1}) we have that $b_0=b^*_0=0$, i.e. 
$\Omega_{\Delta}=\Omega$. Subtituting this equality in Eq. (\ref{Zfin}) and Eq. (\ref{H2}), 
we obtain that
\begin{eqnarray}
\frac{Z}{Z_0}=\frac{1}{(H_I(0))^{1/2}}\,\,
\prod_{\omega\geq 1}\,\frac{1}{\,H_I(\omega)}\,,
\label{Zfinb0}
\end{eqnarray}
where
\begin{eqnarray}
H_I(\omega)&=&\,1\,+\,\frac{(\,g_1^2-g_2^2\,)^2}{(\omega^2+\Omega^2)\,
(\omega^2+\omega^2_0)}\tanh^2\left(\frac{\beta\,\Omega}{2}\right)\,+\nonumber\\
&+&\frac{2\,(g_1^2-g_2^2\,)\,\omega^2\,-\,2\,(\,g_1^2+g_2^2\,)\,\Omega\,\omega_0}
{(\omega^2+\Omega^2)\,
(\omega^2+\omega^2_0)}\, \tanh\left(\frac{\beta\,\Omega}{2}\right)\,.
\label{H21}
\end{eqnarray}
Making the analytic continuation $(i\omega \rightarrow E)$ in $H_I(\omega)$, we
solve the equation $H_I(-i\,E)=0$, which corresponds to the collective spectrum
equation. Solving the equation, we have that
\begin{eqnarray}
2\,E^2&=&\omega_0^2+\Omega^2\,+\,2\,(g_1^2-g_2^2)\,\tanh\left(\frac{\beta\,\Omega}{2}\right)+\nonumber\\
&\pm&\left(\Bigl(\omega_0^2-\Omega^2\Bigr)^2+4\,\Bigl(g_1^2\,(\omega_0+\Omega)^2-g_2^2\,(\omega_0-\Omega)^2\Bigr)
\,\tanh\left(\frac{\beta\,\Omega}{2}\right)\right)^{1/2}\,.
\label{Espb0}
\end{eqnarray}
It is interesting to see, when $\beta =\beta_c$ we find the following roots \cite{aparicio1}
\begin{equation}
E_{\,1}\,=\,0
\label{106}
\end{equation}
and
\begin{equation}
E_{\,2}\,=\,\Biggl(\,\frac{g_{\,1}\,(\Omega\,+\,\omega_{\,0})^{\,2}\,+\,
g_{\,2}\,(\Omega\,-\,\omega_{\,0})^{\,2}}{(g_{\,1}\,+\,g_{\,2})}\,\Biggr)^{\,1/2}\,.
\label{107}
\end{equation}
With Eq. (\ref{Espb0}) we can obtain the collective spectrum
for the two following known cases: the first one, when $g_2=0$, corresponds to the Dicke 
model considering the rotating wave approximation \cite{popov1}. Here we have that
\begin{eqnarray}
2\,E=\omega_0+\Omega\,\pm\,\left(\Bigl(\omega_0-\Omega\Bigr)^2+4\,g_1^2
\,\tanh\left(\frac{\beta\,\Omega}{2}\right)\right)^{1/2}\,.
\label{Espb0g20}
\end{eqnarray}
The second one corresponds to $g_1=g_2=g$. Here we have that
\begin{eqnarray}
2\,E^2=\omega_0^2+\Omega^2\,\pm\,\left(\Bigl(\omega_0^2-\Omega^2\Bigr)^2+16\,g^2\,
\omega_0\,\Omega\,\tanh\left(\frac{\beta\,\Omega}{2}\right)\right)^{1/2}\,.
\label{Espb0g1g2}
\end{eqnarray}
In the case of quantum phase transition, we are in the particular case where $\beta=\infty$. Here,
the collective spectrum corresponds to the Eq. (\ref{Espb0g1g2}) where: 
$\tanh\left(\beta\,\Omega/{2}\right)=1$ \cite{emary2}.

\section{Superradiant phase: $\beta >\beta_c$}

\subsection{Case of $g_1\neq 0$ and $g_2\neq 0$:}
In the superradiant phase $\beta >\beta_c$, in the case of $g_1\neq 0$ and $g_2\neq 0$, we have two maximum points.
Both maximum points contribute equally to the partition function. Therefore $\frac{Z}{Z_0}$, given by Eq. (\ref{Zfin}) 
must be multiplied by a factor $2$. In this case $b_0\neq 0$, i.e. $\Omega_{\Delta}\neq\Omega$. From 
Eq. (\ref{Zfin}), we have that the expresion for $\frac{Z}{Z_0}$ is given by
\begin{eqnarray}
\frac{Z}{Z_0}=2\,e^{N\phi}\frac{1}{(H_{II}(0))^{1/2}}\,\,
\prod_{\omega\geq 1}\,\frac{1}{\,H_{II}(\omega)}\,,
\label{Zfinb1}
\end{eqnarray}
where the factor $\phi$, is defined by
\begin{eqnarray}
\phi\,=\,-\,\frac{\omega_0\,\beta\,(\Omega_{\Delta}^2-\Omega^2)}{4\,(g_1+g_2)^2}+
\ln{\left(\frac{\cosh\left(\frac{\beta\,\Omega_{\Delta}}{2}\right)}
{\cosh\left(\frac{\beta\,\Omega}{2}\right)}\right)}\,.
\label{phi1}
\end{eqnarray}
The function $H_{II}(\omega)$ has the form
\begin{eqnarray}
&&H_{II}(\omega)=\frac{1}{(\omega^2+\Omega_{\Delta}^2)\,(\omega^2+\omega^2_0)}\,\times\nonumber\\
&&\times\left[\,\omega^4+\left(\omega_0^2+\Omega_{\Delta}^2+\frac{2\,(g_1^2-g_2^2)}{(g_1+g_2)^2}\,\omega_0\,\Omega\right)\,
\omega^2+\frac{4\,g_1\,g_2}{(g_1+g_2)^2}\,\omega_0^2\,(\Omega_{\Delta}^2-\Omega^2)\right]\,,
\label{H22}
\end{eqnarray}
and setting $\omega=0$ in Eq. (\ref{H22}), we obtain the expression for $H_{II}(0)$, so that
\begin{eqnarray}
H_{II}(0)=\frac{4\,g_1\,g_2\,(\Omega_{\Delta}^2-\Omega^2)}{(\,g_1+g_2\,)^2\,\Omega_{\Delta}^2}\,.
\label{H022}
\end{eqnarray}
Making the analytic continuation $(i\omega \rightarrow E)$ in $H_{II}(\omega)$ given by Eq. (\ref{H22}), 
we solve the equation $H_{II}(-i\,E)=0$. The set of solutions, $E$, are the collective spectrum in the 
superradiant phase for the case of $g_1\neq 0$ and $g_2\neq 0$. Therefore, solving the equation we have that
\begin{eqnarray}
2\,E^2&=&\omega_0^2+\Omega_{\Delta}^2\,+\,2\,\frac{(\,g_1^2-g_2^2\,)}{(\,g_1+g_2)^2}\,\Omega\,\omega_0\,+\nonumber\\
&\pm&\left[\left(\omega_0^2+\Omega_{\Delta}^2\,+\,2\,\frac{(\,g_1^2-g_2^2\,)}{(\,g_1+g_2)^2}\,\Omega\,\omega_0\right)^2
-\frac{16\,g_1\,g_2}{(\,g_1+g_2)^2}\,\omega_0^2\,\Bigl(\Omega_{\Delta}^2-\Omega^2\Bigr)\right]^{1/2}\,.
\label{Espb1}
\end{eqnarray}
For the particular case of $g_1=g_2=g$, the collective spectrum energy takes the particular form
\begin{eqnarray}
2\,E^2\,=\,\omega_0^2+\Omega_{\Delta}^2\,\pm\,\left((\omega_0^2-\Omega_{\Delta}^2)^2+4\,\omega_0^2\,\Omega^2\right)^{1/2}\,.
\label{Espb1g}
\end{eqnarray}
In the limit of zero temperature, $\beta\rightarrow\infty$, from Eq. (\ref{tra1}) we have that
$\Omega_{\Delta}=4\,g^2/\omega_0$. Consequently, at zero temperature we obtain that \cite{emary2}
\begin{eqnarray}
2\,E^2\,=\,\omega_0^2+\,\frac{16\,g^4}{\omega_0^2}\,\pm\,
\left[\left(\omega_0^2-\frac{16\,g^4}{\omega_0^2}\right)^2+4\,\omega_0^2\,\Omega^2\right]^{1/2}\,.
\label{Espb1g2}
\end{eqnarray}

\subsection{Case of $g_1\neq 0$ and $g_2=0$:}

Now let us study, the case of rotating wave approximation, i. e., the case of $g_1\neq 0$ and $g_2=0$, 
in the superradiant phase. Here, the expression for $\frac{Z}{Z_0}$ is obtained setting $g_2=0$ in Eq. (\ref{ZA2}) 
and Eq. (\ref{fi1}), therefore we have that
\begin{eqnarray}
Z_A=A(N)\int[d\eta(b)]\exp{\left(N\,\Phi_{g_1}(b^*,b)\right)}\,,
\label{ZApopov}
\end{eqnarray}
the function $\Phi_{g_1}(b^*,b)$ is defined by
\begin{eqnarray}
\Phi_{g_1}(b^*,b)=\int_0^{\beta}d\tau\,b^*(\omega)\,(\partial_{\tau}-\omega_0)\,b(\omega)+tr\ln\biggl(1-\,g_1^2\,L_*^{-1}
\,b\,L^{-1}\,b^*\biggr)\,.
\label{fi1g1}
\end{eqnarray}
In last equation, Eq. (\ref{fi1g1}), we can see that the function $\Phi_{g_1}(b^*,b)$ is invariant by 
transformation $b(\tau)\rightarrow\exp{(i\,\theta\,\tau)}\,b(\tau)$ and 
$b^*(\tau)\rightarrow\exp{(-i\,\theta\,\tau)}\,b^*(\tau)$, where $\theta$ 
is an arbitrary factor independent of $\tau$. This continuous invariance is responsible for the appearing 
of Goldstone mode in the system. In order to perform the functional integral given by Eq. (\ref{ZApopov}), 
let us separate the function $b(\tau)$ in the following form
\begin{eqnarray}
b\,(\tau)&=&b_c+b'\,(\tau)\,,\nonumber\\
b^*(\tau)&=&b^*_c+b'^{\,*}(\tau)\,,
\label{bg1}
\end{eqnarray}
where $b_c$ is a constant function, and the fields $b'(\tau)$ and $b'^{\,*}(\tau)$ satisfy the following boundaries 
conditions $b'(0)=b'(\beta)=0$ and $b'^{\,*}(0)=b'^{\,*}(\beta)=0$. Using the representation 
$b_c=\rho\, e^{i\,\phi}$ and $b^*_c=\rho\, e^{-i\,\phi}$ in the functional integral given by Eq. (\ref{ZApopov}) 
and Eq. (\ref{fi1g1}), and after applying the transformation $b'(\tau)\rightarrow e^{i\,\phi}\,b'(\tau)$ and $b'^{\,*}(\tau)\rightarrow e^{-i\,\phi}\,b'^{\,*}(\tau)$, we obtain that
\begin{eqnarray}
Z_A=2\,\pi\,i\,A(N)\,\int_0^{\infty}d\rho^2\,\int[d\eta(b')]\exp{\left(N\,\Phi_{g_1}(\rho,b'^{\,*},b')\right)}\,,
\label{ZApopov1}
\end{eqnarray}
the function $\Phi_{g_1}(\rho,b'^{\,*},b')$ is given by
\begin{eqnarray}
\Phi_{g_1}(\rho,b'^{\,*},b')&=&\int_0^{\beta}d\tau\,\Bigl(\rho+b'^{\,*}(\tau)\Bigr)
\Bigl(\partial_{\tau}-\omega_0\Bigr)\Bigl(\rho+b'(\tau)\Bigr)\,+\nonumber\\
&+&tr\ln\left(1-g_1^2\,L_*^{-1}\,\Bigl(\rho+
b'\Bigr)\,L^{-1}\,\Bigl(\rho+b'^{\,*}\Bigr)\right)\,.
\label{phipopov1}
\end{eqnarray}
In the integral function appearing in Eq. (\ref{ZApopov1}), one variable of integration is $\rho^2$. 
Here we use the steepest descent method in order to analyse the limit $N\rightarrow\infty$, 
we find the stationary point with respect to the variable $\rho^2$. Therefore, the stationary 
point satisfies the following equation $\frac{\delta\,\Phi_{g_1}}{\delta\,(\rho^2)}\Bigr|_{\rho=\rho_0}$ 
with $b'^{\,*}(\tau)=b'(\tau)=0$. In this case the value for $\rho_0$ is the same as $b_0$ defined by 
Eq. (\ref{tra1}) setting $g_2=0$. Let us consider the two first 
leading terms in the functional integral appearing in Eq. (\ref{ZApopov1}), coming from the expansion
of $\Phi_{g_1}(\rho,b'^{\,*},b')$ around the point defined by $\rho_0$ and $b'^{\,*}(\tau)=b'(\tau)=0$, 
giving the maximum for $Re\Bigl(\Phi_{g_1}(\rho,b'^{\,*},b')\Bigr)$. This expansion is given by
\begin{eqnarray}
\Phi_{g_1}(\rho,b'^{\,*},b')&=&\Phi_{g_1}(\rho_0,0,0)+\frac{1}{2}\,\frac{\delta^2\Phi_{g_1}}
{\delta(\rho^2)^2}\Biggr|_{\rho=\rho_0,\,b'=b'^{*}=0}\,\Bigl(\rho^2-\rho_0^2\Bigr)^2\,+\nonumber\\
&+&\frac{1}{2}\int_0^{\beta}d\tau_1\,d\tau_2\,
(b'^{\,*}(\tau_1)\,,\,b'(\tau_1))\,M_{\Phi_{g_1}}
\left(\begin{array}{c}
b'^{\,*}(\tau_2)\\
b'(\tau_2)
\end{array}\right)\,,
\label{fipopov2}
\end{eqnarray}
here the matriz $M_{\Phi_{g_1}}$ is given by
\begin{eqnarray}
\left(\begin{array}{cc}
\frac{\delta^2\Phi_{g_1}}{\delta b'^{*}(\tau_1)\,\delta b'^{*}(\tau_2)}&\frac{\delta^2\Phi_{g_1}}{\delta b'^{*}(\tau_1)\,\delta b'(\tau_2)}\\
\frac{\delta^2\Phi_{g_1}}{\delta b'(\tau_1)\,\delta b'^{*}(\tau_2)}&\frac{\delta^2\Phi_{g_1}}{\delta b'(\tau_1)\,\delta b'(\tau_2)}
\end{array}\right)
\Biggr|_{\rho=\rho_0,\,b'=b'^{*}=0}\,\,.
\label{Mfipopov}
\end{eqnarray}
Using this expansion given by Eq. (\ref{Mfipopov}) to perform functional integral given by Eq. (\ref{ZApopov1}),
we have that
\begin{eqnarray}
Z_A&=&2\,\pi\,i\,\sqrt{N}\,e^{N\phi_{g_1}}\,\int_{-\sqrt{N}\rho^2_0}^{\infty}dy\,e^{\frac{1}{2}\,\frac{\delta^2\Phi_{g_1}}
{\delta(\rho^2)^2}\Bigr|_{\rho=\rho_0,\,b'=b'^{*}=0}\,y^2}\,\times\nonumber\\
&\times&\int[d\eta(b')]\exp\left(\frac{1}{2}\int_0^{\beta}d\tau_1\,d\tau_2\,
(b'^{\,*}(\tau_1)\,,\,b'(\tau_1))\,M_{\Phi_{g_1}}
\left(\begin{array}{c}
b'^{\,*}(\tau_2)\\
b'(\tau_2)
\end{array}\right)\right)\,,
\label{ZApopov2}
\end{eqnarray}
The expression $\phi_{g_1}$ corresponds to the expression of $\phi$ defined in Eq. (\ref{phi1}) taking $g_2=0$. The factor $\sqrt{N}$ appearing in Eq. (\ref{ZApopov2}) 
come from the scaling $\rho^2\rightarrow\rho^2/\sqrt{N}$. For $N\rightarrow\infty$, integrals appearing in Eq. (\ref{ZApopov2})
are Gaussians. We represent the functions $b'(\tau)$ and $b'^{\,*}(\tau)$ in Fourier series, which do not possess the
zero mode, since they satisfy the boundary conditions given by $b'(0)=b'(\beta)=0$ and $b'^{\,*}(0)=b'^{\,*}(\beta)=0$. 
Therefore, performing the funtional integral and substituting in Eq. (\ref{ZA0}), we obtain that
\begin{eqnarray}
\frac{Z}{Z_0}=\sqrt{N}\,e^{N\phi_{g_1}}\,\frac{1}{A_0}\,
\prod_{\omega\geq 1}\,\frac{1}{\,H_{II}(\omega)}\,,
\label{Zfinpopov}
\end{eqnarray}
functions $\phi_{g_1}$ and $H_{II}(\omega)$ are given respectively by Eq. (\ref{phi1}) and Eq. (\ref{H22}) 
setting $g_2=0$, and $A_0$ is given by
\begin{eqnarray}
A_0=\frac{g_1}{\Omega_{\Delta}\,\sqrt{\pi\,\beta\,\omega_0}}\,\left(1-\frac{\beta\,
\Omega_{\Delta}}{\sinh(\beta\Omega_{\Delta})}\right)^{\frac{1}{2}}\,.
\label{A0}
\end{eqnarray}
Making the analytic continuation $(i\omega \rightarrow E)$ in $H_{II}(\omega)$ given by Eq. (\ref{H22})
setting $g_2=0$, the collective spectrum is obtained by solving the equation $H_{II}(-i\,E)=0$. 
So that, we obtain the following spectrum
\begin{eqnarray}
E_1=0\,,
\label{Espg11}
\end{eqnarray}
and
\begin{eqnarray}
E_2^{\,2}=\omega_0^2+\Omega_{\Delta}^2\,+\,2\,\omega_0\,\Omega\,.
\label{Espg12}
\end{eqnarray}
The particular value of the spectrum given by $E_1=0$ in Eq. (\ref{Espg11}) corresponds to the Goldstone mode 
\cite{popov1}.

\subsection{Case of $g_1=0$ and $g_2\neq 0$:}

Now let us study the case of  $g_1=0$ and $g_2\neq 0$, in the superradiant phase. Here, the expression for 
$\frac{Z}{Z_0}$ is obtained setting $g_1=0$ in Eq. (\ref{ZA2}) and Eq. (\ref{fi1}). For this case we have
\begin{eqnarray}
Z_A=A(N)\int[d\eta(b)]\exp{\left(N\,\Phi_{g_2}(b^*,b)\right)}\,,
\label{ZApopovg2}
\end{eqnarray}
the function $\Phi_{g_2}(b^*,b)$ is defined by
\begin{eqnarray}
\Phi_{g_2}(b^*,b)=\int_0^{\beta}d\tau\,b^*(\omega)\,(\partial_{\tau}-\omega_0)\,b(\omega)+tr\ln\biggl(1-\,g_2^2\,L_*^{-1}
\,b^*\,L^{-1}\,b\biggr)\,.
\label{fi1g2}
\end{eqnarray}
In last equation, Eq. (\ref{fi1g2}), we can see that the function $\Phi_{g_2}(b^*,b)$ is invariant by 
transformation $b(\tau)\rightarrow\exp{(i\,\theta\,\tau)}\,b(\tau)$ and 
$b^*(\tau)\rightarrow\exp{(-i\,\theta\,\tau)}\,b^*(\tau)$, where $\theta$ 
is an arbitrary factor independent of $\tau$. This continuous invariance is 
responsible for the appearing of Goldstone mode in the system. Since Eq. (\ref{fi1g2})
is very similar to Eq. (\ref{fi1g1}), we can see that, the calculation to obtain 
$\frac{Z}{Z_0}$ in the case of $g_1=0$ follows the same steps as the calculation performed
to obtain $\frac{Z}{Z_0}$ in the case of rotating wave approximation. Consequently, we have that
\begin{eqnarray}
\frac{Z}{Z_0}=\sqrt{N}\,e^{N\phi_{g_2}}\,\frac{1}{A_0}\,
\prod_{\omega\geq 1}\,\frac{1}{\,H_{II}(\omega)}\,,
\label{Zfinpopovg2}
\end{eqnarray}
where $H_{II}(\omega)$ is given by Eq. (\ref{H22}) setting $g_1=0$, the value $\phi_{g_2}$ 
corresponds to the expression for $\phi$ defined in Eq. (\ref{phi1}) taking $g_1=0$, and $A_0$ is given by
\begin{eqnarray}
A_0=\frac{g_2}{\Omega_{\Delta}\,\sqrt{\pi\,\beta\,\omega_0}}\,\left(1-\frac{\beta\,
\Omega_{\Delta}}{\sinh(\beta\Omega_{\Delta})}\right)^{\frac{1}{2}}\,.
\label{A0g2}
\end{eqnarray}
Making the analytic continuation $(i\omega \rightarrow E)$ in $H_{II}(\omega)$ given by Eq. (\ref{H22})
setting $g_1=0$, 
the collective spectrum is obtained by solving the equation $H_{II}(-i\,E)=0$. So that, we obtain the
following spectrum
\begin{eqnarray}
E_1=0\,,
\label{Espg21}
\end{eqnarray}
and
\begin{eqnarray}
E_2^{\,2}=\omega_0^2+\Omega_{\Delta}^2\,-\,2\,\omega_0\,\Omega\,.
\label{Espg22}
\end{eqnarray}
The particular value of the spectrum given by $E_1=0$ in Eq. (\ref{Espg21}) corresponds to the Goldstone mode.

\section{Summary}

\quad $\,\,$
In this paper, using the path integral approach and functional methods, in
the thermodynamic limit $N\rightarrow\infty$, we find the 
asymptotic behaviour of the partition function and collective spectrum  
of the full Dicke model in the normal and superradiant phase. In our study we 
distinguish three particular cases.
The first one corresponds to the case of rotating wave approximation, 
$g_1\neq 0$ and $g_2=0$, in this case the model has a continuous symmetry, which is associated
to the conservation of the sum of the number excitation of the $N$ atoms with the number 
excitation of the boson field. The second case corresponds to the model with $g_1=0$ and $g_2\neq 0$,
in this case the model has a continuous symmetry, which is associated to the conservation 
of the difference between the number excitation of the $N$ atoms and the number 
excitation of the boson field. The last one, corresponds to the case of 
$g_1\neq 0$ and $g_2\neq 0$, which the model has a discrete symmetry. The phase transition in each case is 
related to the spontaneous breaking of their respective symmetry. In the case of rotating 
wave approximation, and also in the case of $g_1=0$ and $g_2\neq 0$, the collective spectrum has a zero energy
value, corresponding to the Goldstone mode associated to the continuous symmetry breaking for these cases.

\section{Acknowledgements}

MAA was supported by FAPESP, and BMP was partially supported by CNPq.

\end{document}